\documentclass[twocolumn,showpacs,prl]{revtex4}

\usepackage[dvips]{graphicx}
\usepackage{dcolumn}
\usepackage{bm}

\begin{document}

\title{Electrically detected electron spin resonance in a high mobility silicon quantum well}

\author{Junya Matsunami}

\author{Mitsuaki Ooya}

\author{Tohru Okamoto}

\affiliation{Department of Physics, University of Tokyo, 7-3-1 Hongo, Bunkyo-ku, Tokyo 113-0033, Japan}

\date{today}

\begin{abstract}

The resistivity change due to electron spin resonance (ESR) absorption is investigated in a high-mobility two-dimensional electron system formed in a Si/SiGe heterostructure.
Results for a specific Landau level configuration demonstrate that the primary cause of the ESR signal is a reduction of the spin polarization, not the effect of electron heating.
The longitudinal spin relaxation time $T_1$ is obtained to be of the order of 1~ms in an in-plane magnetic field of 3.55~T.
The suppression of the effect of the Rashba fields due to high-frequency spin precession explains the very long $T_1$.

\end{abstract}


\maketitle

Electron spin in semiconductor devices has recently attracted much attention in the context of the quantum computation and spintronics application \cite{Zutic2004}. 
Electron spin resonance (ESR) is a promising technique to manipulate spins directly, and electrical detection via a change in resistivity is preferable for the read out of the local spin states.
However, the origin of the resistivity change due to ESR absorption has not been established. 
In previous works performed on GaAs two-dimensional electron systems (2DESs) in high magnetic fields \cite{Stein1983,Dobers1988,Meisels2000,Olshanetsky2003}, the sign of the change $\Delta \rho_{xx}$ in longitudinal resistivity $\rho_{xx}$ was found to be the same as that of the derivative of $\rho_{xx}$ with respect to temperature $T$ and the effect of electron heating has not been excluded.

For fabricating spintronics devices, silicon appears to be a very suitable host material.
Addition to the compatibility with the fabrication technology of integrated circuits, it has the advantage of long electron spin relaxation times due to weak spin-orbit interactions and poor electron-nuclear spin (hyperfine) coupling.
Recently, low magnetic field ($\simeq 0.3~{\rm T}$) ESR measurements have been performed in silicon 2DESs formed in Si/SiGe heterostructures \cite{Graeff1999,Wilamowski2002,Wilamowski2004,Tyryshkin2005}.
Both the longitudinal spin relaxation time $T_1$ and the transverse spin relaxation time $T_2$ were measured to be of the order of $1~\mu$s and were discussed in terms of the D'yakonov-Perel' spin relaxation mechanism \cite{Dyakonov1971} due to the Rashba fields \cite{Bychkov1984}.

In this Letter, we report electrically detected ESR measurements on a high-mobility silicon 2DES in a magnetic field of 3.55~T. 
Negative $\Delta \rho_{xx}$ clearly observed for a specific Landau level (LL) configuration in a tilted magnetic field demonstrates that the ESR signal is mainly caused by a reduction of the spin polarization $P$, not by the electron heating effect.
For a quantitative discussion, measurements were also performed in the magnetic field $B_\parallel$ applied parallel to the 2D plane.
The $T$-dependence of the ESR signal is well reproduced with $T$-independent spin relaxation times.
From the amplitude of the ESR signal, $T_1$ is estimated to be of the order of 1~ms whereas $T_2$ deduced from the linewidth is about 10~ns.
The enhancement of $T_1$ is explained by the suppression of the effect of the Rashba fields due to high-frequency spin precession.

We used a heterostructure sample with a 20-nm-thick strained Si channel sandwiched between relaxed $\mathrm{Si}_{0.8}\mathrm{Ge}_{0.2}$ layers \cite{Yutani1996}. 
The 2D electron concentration $N_s$ can be controlled by varying the bias voltage of a $p$-type Si~(001) substrate 2.1~$\mu \mathrm{m}$ below the channel. 
The 2DES has a high mobility of $48~\mathrm{m^2/V~s}$ at $N_s=2.1\times 10^{15}~\mathrm{m^{-2}}$ (at the zero substrate bias voltage) and $T=0.31~{\text K}$. 
Four-probe ac resistivity measurements (137~Hz) were performed for a $600\times50~\mu \mathrm{m}^2$ Hall bar sample. 
The ESR signal $\Delta \rho_{xx}$ was observed during the magnetic field sweep in the presence of 100~GHz millimeter wave radiation. 
The sample was mounted inside an oversized waveguide with an 8~mm bore inserted into a pumped ${}^{3}{\rm He}$ refrigerator.
While the angle of the 2D plane with respect to the magnetic field generated by a Helmholtz magnet can be controlled by rotating the sample, the direction of the source-drain current was fixed perpendicular to that of the magnetic field. 

\begin{figure}[tb]
\begin{center}
\includegraphics[width=8cm]{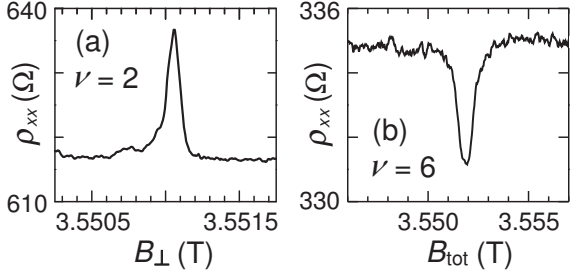}
\caption{\label{fig1}
ESR signals with a millimeter-wave output power of 8~mW. (a) Data for $\nu=2$ in a perpendicular magnetic field ($ B_{\text {tot}}/B_{\bot}=1$) at $N_s=1.74\times 10^{15}~\mathrm{m^{-2}}$ and $T=2.0~{\text K}$. (b) Data for $\nu=6$ and $ B_{\text {tot}}/B_{\bot}=3.36$ at $N_s=1.58\times 10^{15}~\mathrm{m^{-2}}$ and $T=1.1~{\text K}$.
}
\end{center}
\end{figure}

We first discuss ESR in the quantum Hall systems where the electronic states and transport properties are well understood. 
In the previous measurements on GaAs 2DESs at $\nu={\rm odd}$, positive $\Delta \rho_{xx}$ due to ESR have been reported \cite{Stein1983,Olshanetsky2003}. 
Shown in Fig.~\ref{fig1}(a) is a positive ESR signal also observed at $\nu=2$ in our silicon 2DES with the twofold valley degeneracy. 
In these measurements, the Zeeman splitting $E_{\rm Zeeman}$ is smaller than the cyclotron gap $\hbar \omega _c$, where $\omega _c$ is the cyclotron frequency.
The chemical potential $\mu$ lies between the spin-up and spin-down LLs with the same orbital index $n$. 
Both electron heating and spin-flips directly induced by ESR increase the number of excited carriers in the LLs and thus lead to the positive $\Delta \rho _{xx}$.

\begin{figure}[tb]
\begin{center}
\includegraphics[width=7.5cm]{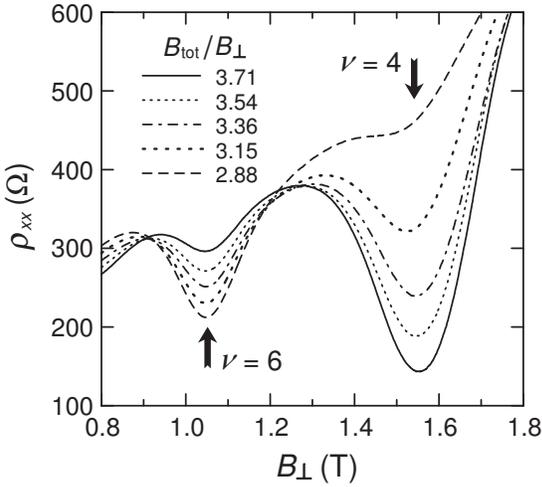}
\caption{\label{fig2}
Shubnikov-de Haas oscillations for different values of $ B_{\text {tot}}/B_{\bot}$ at $N_s=1.58\times 10^{15}~\mathrm{m^{-2}}$ and $T=0.4~{\text K}$. 
Note that the twofold valley degeneracy remains in 2DESs formed in Si~(001) quantum wells. 
}
\end{center}
\end{figure}

\begin{figure}[tb]
\begin{center}
\includegraphics[width=8.3cm]{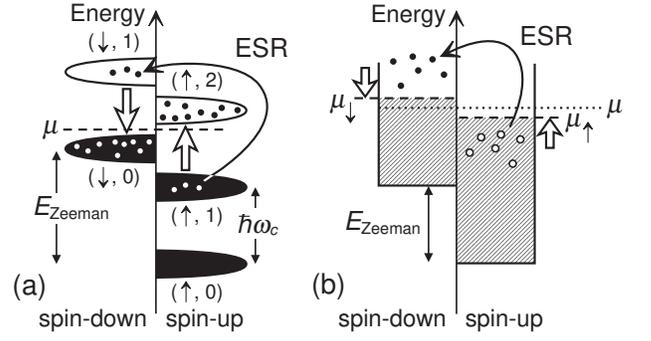}
\caption{\label{fig3}
(a) Landau level configuration and ESR-induced carrier dynamics. 
The LLs are labeled by their spin orientation and orbital index. 
(b) ESR-induced carrier dynamics under an in-plane magnetic field. Photoexcited carriers immediately relax their energy to the lattice and leads to a downward (upward) shift of the chemical potential $\mu_\uparrow$ ($\mu_\downarrow$) for spin-up (spin-down) electrons from the equilibrium chemical potential $\mu$. A broadening of the Fermi distribution function at finite lattice temperatures is not drawn for clarity.
}
\end{center}
\end{figure}

To determine the mechanism of $\Delta \rho_{xx}$ induced by ESR, we made a specific LL configuration by adjusting the angle of the 2D plane with respect to the magnetic field. 
The total strength $B_{\rm tot}$ and the perpendicular component $B_\perp$ of the magnetic field are proportional to $E_{\rm Zeeman}$ and $\hbar \omega_c$, respectively.
Figure~\ref{fig2} shows Shubnikov-de Haas oscillations for different values of $B_{\rm tot}/B_\perp$. 
The minimal value of $\rho_{xx}$ at $\nu =6 $ increases with $B_{\rm tot}$ while that at $\nu =4$ decreases. 
This indicates that the energy separation between the LLs just above and below $\mu$ decreases with increasing $E_{\rm Zeeman}$ at $\nu=6$ for this region of $B_{\rm tot}/B_\perp$. 
We consider that $\mu$ lies between the spin-down LL with $n=0$ (LL$(\downarrow, 0)$) and the spin-up LL with $n=2$ (LL$(\uparrow, 2)$) as illustrated in Fig.~\ref{fig3}(a). 


In contrast to the ESR measurements for $E_{\rm Zeeman} < \hbar \omega_c$, we observed negative $\Delta\rho_{xx}$ at $\nu=6$ for $B_{\rm tot}/B_\perp = 3.36$ as shown in Fig.~\ref{fig1}(b). 
Since the $T$-dependence of $\rho_{xx}$ at the minimum is positive, electron heating cannot be the cause of the negative ESR signal. 
In order to explain it, we consider the relaxation process of photoexcited carriers as illustrated in Fig.~\ref{fig3}(a). 
Since the orbital index $n$ does not change during the spin-flip, the photoexcitation occurs mainly from the filled LL$(\uparrow, 1)$ to the empty LL$(\downarrow, 1)$. 
Although photoexcited electrons in the LL$(\downarrow, 1)$ and holes in the LL$(\uparrow, 1)$ can increase $\rho_{xx}$, they are expected to immediately relax to the LL$(\downarrow, 0)$ and the LL$(\uparrow, 2)$, respectively, due to electron-phonon interactions.
Since the electron-lattice relaxation time $\tau_{\it e-l}$ is much shorter than the longitudinal spin relaxation time $T_1$ as described later, the total number of carriers is reduced by the recombination of photoexcited carriers with thermally activated carriers.
In this LL configuration, the sign of $\Delta \rho_{xx}$ due to a reduction of the spin polarization $P$ is negative.

In order to give a quantitative discussion, we now extend the study to the in-plane magnetic field configuration.
It is well-known that silicon 2DESs show strong positive dependence of resistivity $\rho$ on $B_\parallel$ \cite{Abrahams2001}.
Since $B_\parallel$ does not directly affect the 2D motion of carriers, the $B_\parallel$-dependence of $\rho$ is attributed to the spin polarization \cite{Okamoto1999}.
While there is no consensus on the mechanism of the increase in $\rho$ with $P$ \cite{Phillips1998,Dolgopolov2000,Herbut2001,Zala2002,Spivak2003,Sarma2005}, we here use $\rho$ as a measure of $P$.
The positive dependence of $\rho$ on $P$ for a given temperature can be empirically derived from the magnetoresistance curve in the absence of radiation since $P$ is simply calculated as a function of $B_{\parallel}$ and $T$ from the Fermi distribution of the 2DES \cite{Tejun}.
Typical data for $\rho$ vs $B_\parallel$ at low $T$ in the present sample have been presented in Ref.~\cite{Okamoto2004}.
The critical magnetic field $B_c$ for the full spin polarization at $T=0$ is higher than $B_\parallel=3.55$~T at which the ESR measurements were performed.

The sign of the resistivity change $\Delta \rho$ due to ESR was found to be negative in the experimental range of $T$ and $N_s$.
Electron heating cannot cause the negative $\Delta \rho$ because $\partial \rho / \partial T$ is positive \cite{Okamoto2004}.
The photoexcited electrons and holes are expected to relax their energy to the lattice immediately without spin relaxation, as in the specific LL configuration discussed above (see Fig.~3(b)).
From the relationship between the electron temperature and the Joule heating energy, $\tau_{\it e-l}$ is deduced to be of the order of or less than 1~ns \cite{Ooya}.
Under continuous wave excitation, the slow spin relaxation allows a downward (upward) shift of the chemical potential $\mu_\uparrow$ ($\mu_\downarrow$) for spin-up (spin-down) electrons from the equilibrium chemical potential $\mu$.
This leads to the reduction of $P$ and the negative $\Delta \rho$
while the origin of negative $\partial \rho / \partial P$ is not the same as that in the quantum Hall regime.

In Fig.~4 the peak value $\Delta\rho_{\rm peak}$ of the negative ESR signal is plotted against $T$ at $N_s =1.43 \times 10^{15}~{\rm m}^{-2}$ with a constant millimeter-wave output power.
The amplitude of $\Delta\rho_{\rm peak}$ decreases with increasing $T$ while the full width at half the maximum of the ESR signal has a $T$-independent value of 1.2~mT.
Neglecting the resonant electron heating effect on $\rho$, $\Delta \rho$ is directly related to the change $\Delta P$ in $P$ as $\Delta \rho = (\partial \rho / \partial P) \Delta P$.
The partial derivative $\partial \rho / \partial P$ can be evaluated from experimental data on $\partial \rho/\partial B_{\parallel}$ in the absence of radiation \cite{rhoB} with calculated $\partial P/\partial B_{\parallel}$.
The dashed line in Fig.~4 represents the calculation of $\Delta \rho_{\rm peak}$ assuming a $T$-independent value of $\Delta P/P = -0.08$.
The general agreement with the experimental data indicates that the $T$-dependence of $\Delta P/P$ at the peak is weak.
The $T$-dependence of $\Delta \rho_{\rm peak}$ is attributed to decreases in $\partial \rho/\partial P$ and $P$ with increasing $T$.

To obtain the spin relaxation times from $\Delta P/P$, we use the solution of the Bloch equations for continuous wave radiation \cite{Slichter}.
Substituting Eq.~(2.90) into Eq.~(2.86a) in Ref.~\cite{Slichter} yields
\begin{eqnarray}
\frac{\Delta P}{P} = -\frac{(\gamma B_1)^2 T_1 T_2}{4} 
\frac{1}{1+(\omega - \gamma B_\parallel)^2 T_2 {}^2}
,
\end{eqnarray}
where $\gamma$ is the electron gyromagnetic ratio, $B_1$ is the amplitude of the component of the oscillating magnetic field perpendicular to the static magnetic field, and $\omega$ is its frequency ($=2\pi \times 10^{11}~{\rm s}^{-1}$), respectively \cite{B1def}.
As expected from Eq.~(1), the observed ESR signal has a Lorentzian lineshape. 
We obtain an almost $T$- and $N_s$-independent value of $T_2\simeq 10~{\rm ns}$ as shown in Fig.~5(a) from the linewidth with an accuracy of 6~\%. 
The weak $T$-dependence of the peak value of $\Delta P/P$ indicates that $T_1$ is nearly $T$-independent in the experimental range while it has an error of 30~\% arising from that in $\partial \rho / \partial P$ \cite{rhoB}.
The value of $B_1$ can be roughly estimated from its relationship $B_1 \sim E_1 n_{\rm Si}/c$ to the amplitude of the oscillating electric field $E_1$, where $n_{\rm Si}$ is the refractive index of silicon and $c$ is the speed of light \cite{B1}.
From electron cyclotron resonance absorption measurements made in a perpendicular static magnetic field, we obtain $E_1 = 67~{\rm V/m}$ for a millimeter-wave output power of 19~mW \cite{Ooya}.
In Fig.~5(b), $T_1$ estimated from Eq. (1) by putting $B_1 = E_1 n_{\rm Si}/c$~$(=0.8~\mu {\rm T})$ is shown against $N_s$ \cite{PD}. 
The obtained values of $T_1$ ($1 \sim 3~{\rm ms}$) are five orders of magnitude longer than $T_2 \simeq 10~{\rm ns}$, whereas both $T_1$ and $T_2$ were reported to be of the order of $\mu$s in previous studies on lower-mobility Si/SiGe samples in low magnetic fields ($B \simeq 0.3~{\rm T}$) \cite{Graeff1999,Wilamowski2004,Tyryshkin2005}. 

\begin{figure}[t]
\begin{center}
\includegraphics[width=6.5cm]{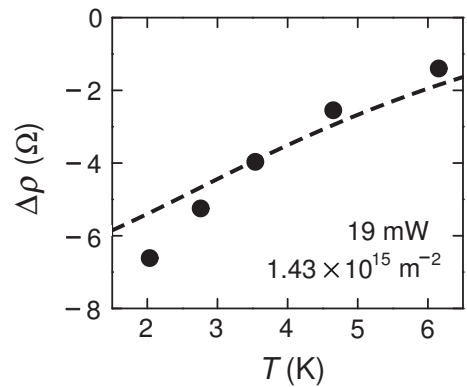}
\caption{\label{fig4} 
Peak value of the ESR signal in the in-plane magnetic field as a function of $T$ at $N_s=1.43\times 10^{15}~\mathrm{m^{-2}}$ with a millimeter-wave output power of 19~mW.
The dashed line is calculated assuming $\Delta P/P=-0.08$.
}
\end{center}
\end{figure}
\begin{figure}[t]
\begin{center}
\includegraphics[width=7cm]{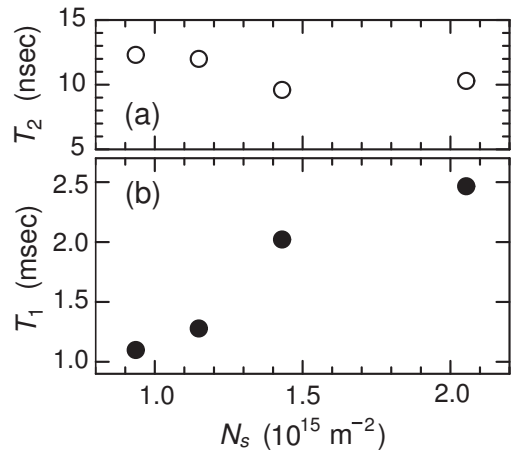}
\caption{\label{fig5} 
Electron spin relaxation times, (a) $T_2$ and (b) $T_1$, in the in-plane magnetic field of 3.55~T as functions of $N_s$.
}
\end{center}
\end{figure}

Spin relaxation in silicon 2DESs is considered to be dominated by the D'yakonov-Perel' mechanism \cite{Dyakonov1971} via the Rashba fields, whose direction depends on the electron momentum direction \cite{Bychkov1984}. 
The spin relaxation rates in the in-plane magnetic field configuration are given by \cite{Slichter,Tahan2005}
\begin{eqnarray}
\frac{1}{T_1} &=& \frac{1}{2} \gamma ^2 B _{{\rm R}} {}^2 \tau _c \ \frac{1}{1 + \omega _L {}^2 \tau _c {}^2} \label{eq:T1}, \\
\frac{1}{T_2} &=& \frac{1}{2} \gamma ^2 B _{{\rm R}} {}^2 \tau _c + \frac{1}{2T_1} \label{eq:T2}
,
\end{eqnarray}
where $B _{{\rm R}}$ is the amplitude of the Rashba field, $\tau_c$ is the correlation time of the field fluctuations, and $\omega _L$ ($ = \gamma B_\parallel$) is the Larmor frequency.
The factor $1/(1 + \omega_L {}^2 \tau_c {}^2)$ in Eq.~(2) represents the suppression of the effect of the Rashba fields due to high-frequency spin precession. 
The ratio is expected to be $T_1/T_2 \sim \omega _L {}^2 \tau _c {}^2$ for $\omega_L \tau_c \gg 1$ while $T_1/T_2 \sim 1$ for $\omega_L \tau_c \lesssim 1$.
This is consistent with the fact that large values of $T_1/T_2$ are obtained in our high-mobility sample located in a high magnetic field.
The results demonstrate a possible way to attain long $T_1$, which is one of the crucial requirements for the semiconductor spintronics application \cite{Wilamowski2004,Tyryshkin2005}.

It seems reasonable to assume that $\tau_c$ is the same as the momentum scattering time $\tau_m$ ($=11-44~{\rm ps}$) determined from $\rho$ ($=1.05-0.10~{\rm k}\Omega$ at $B_\parallel=3.55~{\rm T}$).
However, $\omega_L {}^2 \tau_m {}^2$ ($=48-760$) is much less than the observed $T_1/T_2$.
The longitudinal spin relaxation rate ($=T_1 {}^{-1}$) is related to the Fourier transform of the correlation function $G(t)$ of the Rashba field \cite{Wilamowski2004,Slichter,Tahan2005}.
A single exponential decay of $G(t) \propto \exp(-|t|/\tau_c)$ leads to the Fourier transform of
\begin{eqnarray}
\int_{-\infty}^{\infty} \exp(i \omega_L t) G(t) d t \propto \frac{\tau_c}{1+\omega _L {}^2 \tau _c {}^2}
\label{eq:INTEG}
,
\end{eqnarray}
which appears in Eq.~(\ref{eq:T1}).
This integral is very sensitive to the shape of $G(t)$ particularly for  $\omega_L \tau_c \gg 1$ \cite{Fourier}.
One of the important scattering mechanisms in the present Si/SiGe heterostructure sample is the Coulomb scattering by remote ionized dopants \cite{Yutani1996,YutaniDoc}.
It is unlikely that $G(t)$ for scattering in long-range potential fluctuations can be described exactly by a simple expression with a single parameter.
The deviation of $G(t)$ from a single exponential may cause the discrepancy between $T_1/T_2$ and $\omega_L {}^2 \tau_c {}^2$.

In the presence of high $B_\perp$, the cyclotron motion of electrons is expected to cause averaging of the Rashba fields \cite{Wilamowski2004}, and Eqs.~(2) and (3) should be modified.
This effect can increase the spin relaxation times from the values at $B_{\perp}=0$.
From the linewidths of the ESR signals shown in Fig.~1, enhanced $T_2$ ($=110$~ns) is actually obtained for $\nu=2$ ($B_\perp=3.55$~T) while $T_2=13$~ns for $\nu=6$ ($B_\perp=1.06$~T) is comparable to that in the in-plane magnetic field ($B_\perp=0$).
Further studies are required for quantitative analysis at high $B_\perp$.

In summary, we have investigated the resistivity change due to ESR absorption in a high-mobility silicon 2DES in a magnetic field of 3.55~T. 
Negative $\Delta \rho_{xx}$ observed for a specific LL configuration clearly shows that $\Delta \rho_{xx}$ is mainly caused by a reduction of $P$, not by the electron heating effect.
In the in-plane magnetic field, negative $\Delta \rho$ was used for quantitative analysis.
It is estimated that $T_1$ is of the order of 1~ms, which is five orders of magnitude longer than $T_2$.
This is explained by the suppression of the effect of the Rashba fields due to high-frequency spin precession.

We thank Y. Shiraki for providing us with the Si/SiGe sample and S. Yamamoto for helpful advice on the use of the millimeter wave system.
This work is supported in part by Grants-in-Aid for Scientific Research from the Ministry of Education, Science, Sports, and Culture of Japan, and the Sumitomo Foundation.


\begin{references}
\bibitem{B1def}
The factor of 1/4 in Eq.~(1) comes from the definition of $B_1$. In Ref.~\cite{Slichter} the millimeter wave is taken to be circularly polarized whereas we take it to be linearly polarized.
\bibitem{B1}
Considering multiple reflection in the sample, $B_1 c / E_1 n_{\rm Si}$ is evaluated within the range of 0.5 to 1.9.
\bibitem{PD}
Similar results are obtained from measurements with a millimeter-wave output power of 8~mW.
\bibitem{Fourier}
For example, if $G(t) \propto \exp(-t^2 / \tau_c {}^2)$ is used, the integral is proportional to $\tau_{c} \exp(-\omega_L {}^2 \tau_c {}^2 /4)$.
\end{references}
\end{document}